\documentstyle[twoside,fleqn,espcrc2,psfig,epsfig]{article}

\title{Quantum chaos and regularity in $\Phi^4$ theory }

\author{Helmut Kr\"oger\address{D\'epartement de Physique, Universit\'e Laval, Qu\'ebec, Qu\'ebec G1K 7P4, Canada}, 
       Xiang-Qian Luo\address{Department of Physics, Zhongshan University, Guangzhou 510275, China},
       Harald Markum\address{Atominstitut, Technische Universit\"at Wien, A-1040 Wien, Austria},
       Rainer Pullirsch$^{\rm c}$}
       
\begin{document}

\begin{abstract}
We check the eigenvalue spectrum of the $\Phi^{4}_{1+1}$ Hamiltonian against Poisson
or Wigner behavior predicted from random matrix theory. We discuss random matrix
theory as a tool to discriminate the validity of a model Hamiltonian compared to
an analytically solvable Hamiltonian or experimental data.
\end{abstract}

\maketitle

\section{Motivation}

The fluctuation properties of the eigenvalues of 
a Hamiltonian give much insight into the dynamics of the underlying system.
In particular, the so-called nearest-neighbor spacing distribution $P(s)$, i.e.,
the distribution of spacings $s$ between adjacent eigenvalues plays an important 
role. According to the Bohigas-Giannoni-Schmit conjecture \cite{Bohi84}, quantum systems whose
classical counterparts are chaotic have a $P(s)$ given by random matrix theory (RMT)
whereas systems whose classical counterparts are integrable obey a Poisson distribution,
$P(s)=e^{-s}$.  In this sense, the form of $P(s)$ characterizes quantum chaos.

Today we know a number of physical systems where Wigner and Poisson behavior is
observed~\cite{Guhr}. Here we test a one-dimensional chain of $N_{s}$ coupled
harmonic oscillators, with anharmonic perturbation~\cite{selig}.
Its Euclidean action is given by
\begin{eqnarray}
\label{action_osc}
S = \int dt \sum_{n=1}^{N_{s}} ~ \frac{1}{2} \dot{\phi}_{n}^{2} 
+ \frac{\Omega^{2}}{2} (\phi_{n+1} - \phi_{n})^{2} \nonumber \\
+ \frac{\Omega_{0}^{2}}{2} \phi_{n}^{2} 
+ \frac{\lambda}{2} \phi_{n}^{4} ~ .
\end{eqnarray}
In the continuum formulation it corresponds to the scalar $\Phi^{4}_{1+1}$ model,
\begin{eqnarray}
S = \int dt \int dx ~ 
\frac{1}{2} (\frac{\partial \Phi}{\partial t})^{2} 
+ \frac{1}{2} (\nabla_{x} \Phi)^{2}  \nonumber \\
+ \frac{m^{2}}{2} \Phi^{2} 
+ \frac{g}{4!} \Phi^{4} ~ .
\end{eqnarray}
Introducing a space-time lattice with lattice spacing $a_{s}$ and $a_{t}$, this action becomes
\begin{eqnarray}
\label{action_scalar}
S & = & \sum_{n=1}^{N_{s}} \sum_{k=0}^{N_{t}-1} a_{t} a_{s} \nonumber \\
 &\phantom{\times}& \left[
\frac{1}{2} \left( \frac{ \Phi(x_{n},t_{k+1}) - \Phi(x_{n},t_{k}) }{a_{t}} \right)^{2}  \right.
\nonumber \\
 &+&  \frac{1}{2} \left( \frac{ \Phi(x_{n+1},t_{k}) - \Phi(x_{n},t_{k}) }{a_{s}} \right)^{2}
\nonumber \\
&+& \left. \frac{m^{2}}{2} \Phi^{2}(x_{n},t_{k}) 
+ \frac{g}{4!} \Phi^{4}(x_{n},t_{k}) \right] ~ .
\end{eqnarray}
The actions given by Eq.(\ref{action_scalar}) and Eq.(\ref{action_osc})
can be identified by posing $\phi=\sqrt{a_s}\Phi$, $\Omega=1/a_s$,
$\Omega_0=m$, and $\lambda/2=g/4!$.

For the underlying real and symmetric matrix one expects a correspondence to the
Gaussian orthogonal ensemble (GOE).
The RMT result for $P(s)$ is quite complicated; it can be expressed in terms of
so-called prolate spheroidal functions, see Ref.~\cite{Meht91} where $P(s)$ has
also been tabulated.  A very good approximation to $P(s)$ is provided by \begin{equation}
  \label{eq1}
  P(s)=\frac{\pi}{2}s e^{-\frac{\pi}{4}s^2}
\end{equation}
which is the Wigner surmise for the GOE.

\begin{figure*}[hp]
\centerline{\hspace*{45mm} analytic \hspace*{62mm} stochastic}\vspace*{-5mm}
\centerline{{\hspace*{-5mm}\psfig{figure=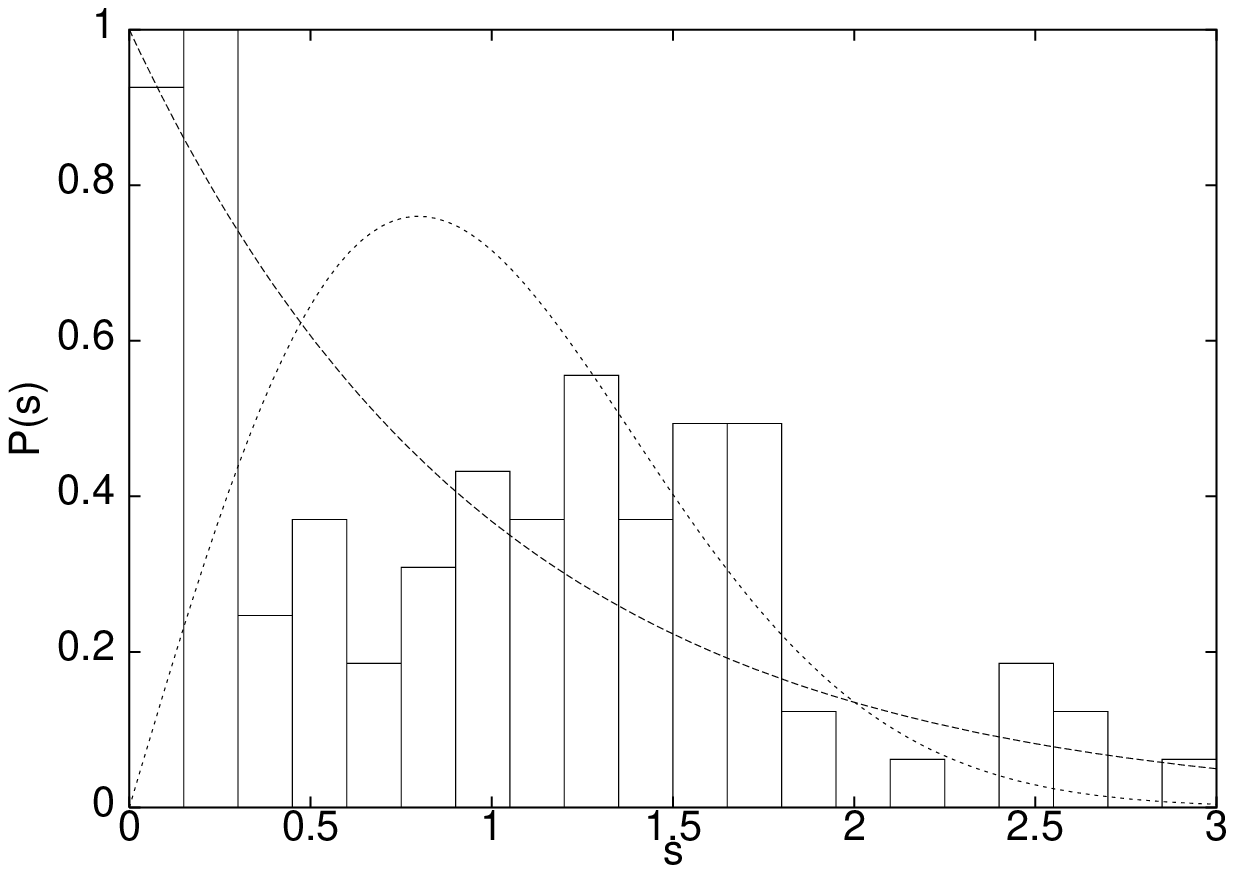,width=5cm}}\hspace{25mm}
{\psfig{figure=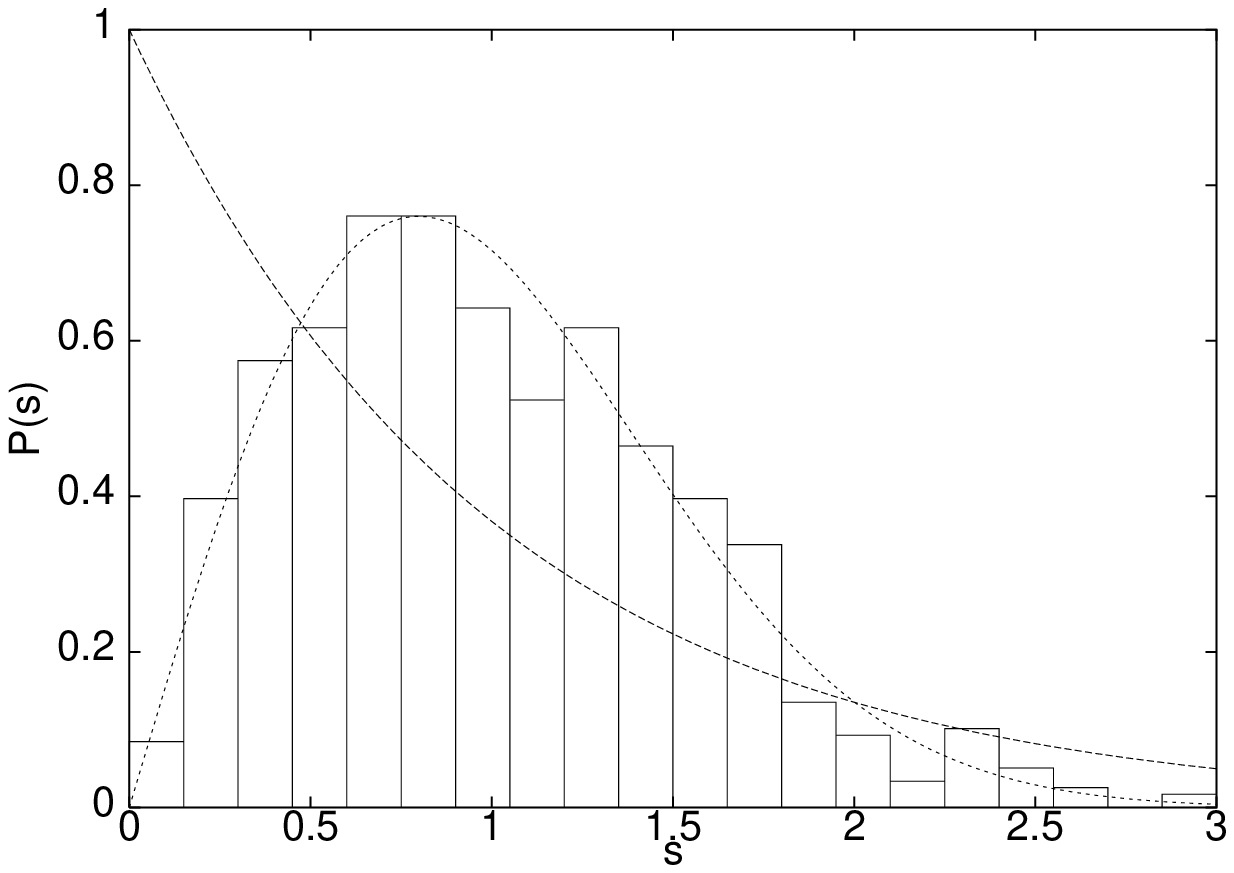,width=5cm}}}
\vspace*{0mm}
\caption{Nearest-neighbor spacing distribution $P(s)$ for a chain with
$N_s =9 $ harmonic oscillators from the analytical spectrum (left plot)
and from the stochastic method (right plot). The Poisson and the Wigner
distribution for the GOE are inserted.
\vspace*{15mm}
\label{fig1}
}

\centerline{\hspace*{47.5mm} $N_s = 3$ \hspace*{63.5mm} $N_s = 5$}\vspace*{-5mm}
\centerline{{\hspace*{-5mm}\psfig{figure=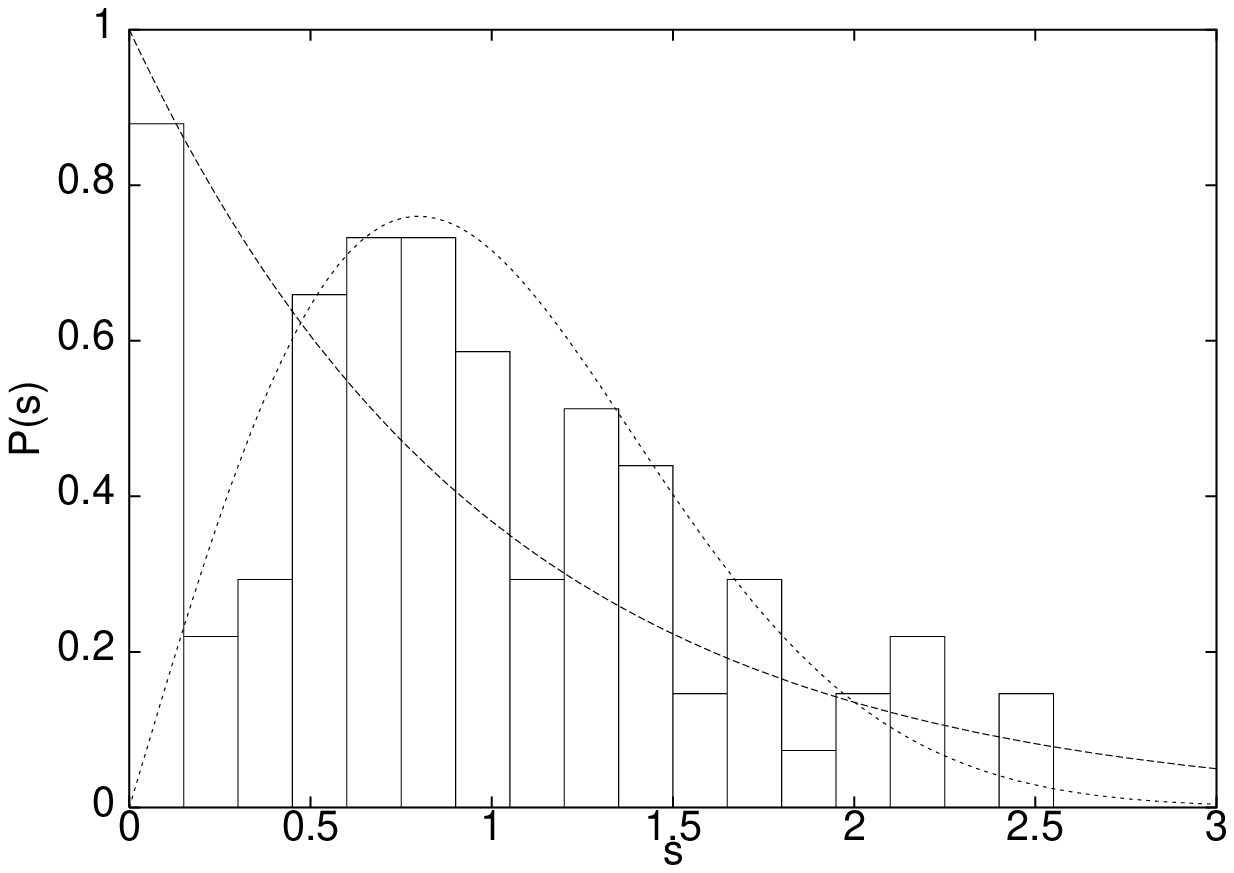,width=5cm}}\hspace{25mm}
{\psfig{figure=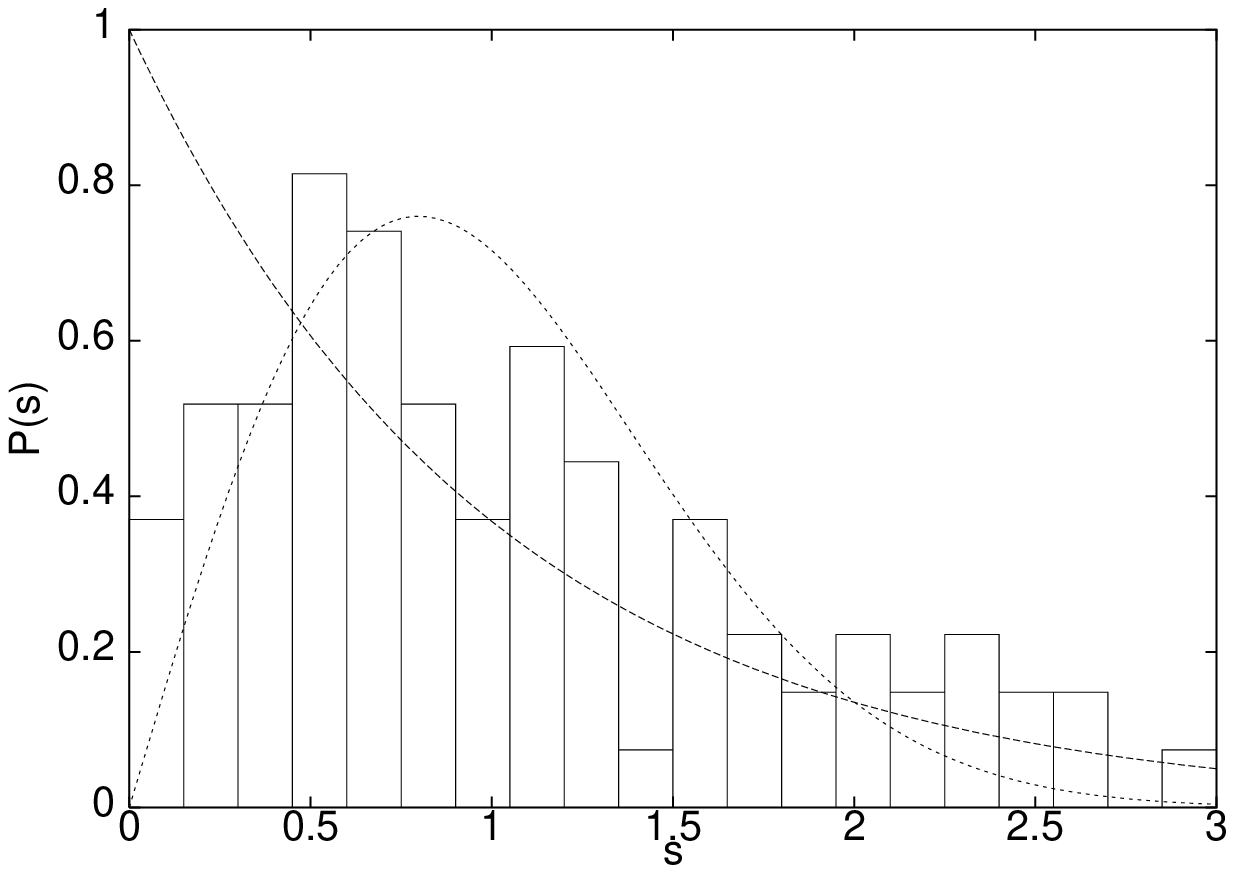,width=5cm}}}
\vspace*{5mm}
\centerline{\hspace*{47.5mm} $N_s = 7$ \hspace*{63.5mm} $N_s = 9$}\vspace*{-5mm}
\centerline{{\hspace*{-5mm}\psfig{figure=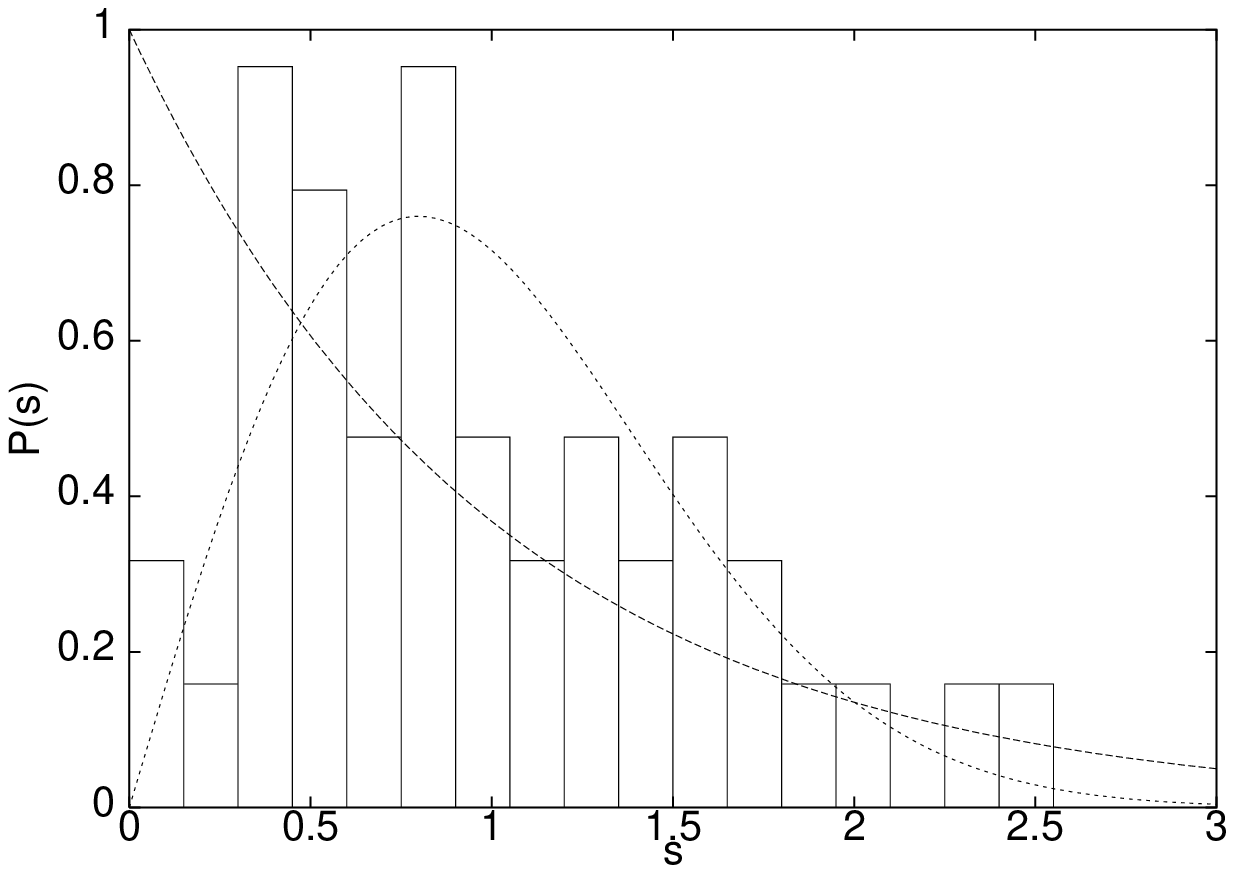,width=5cm}}\hspace{25mm}
{\psfig{figure=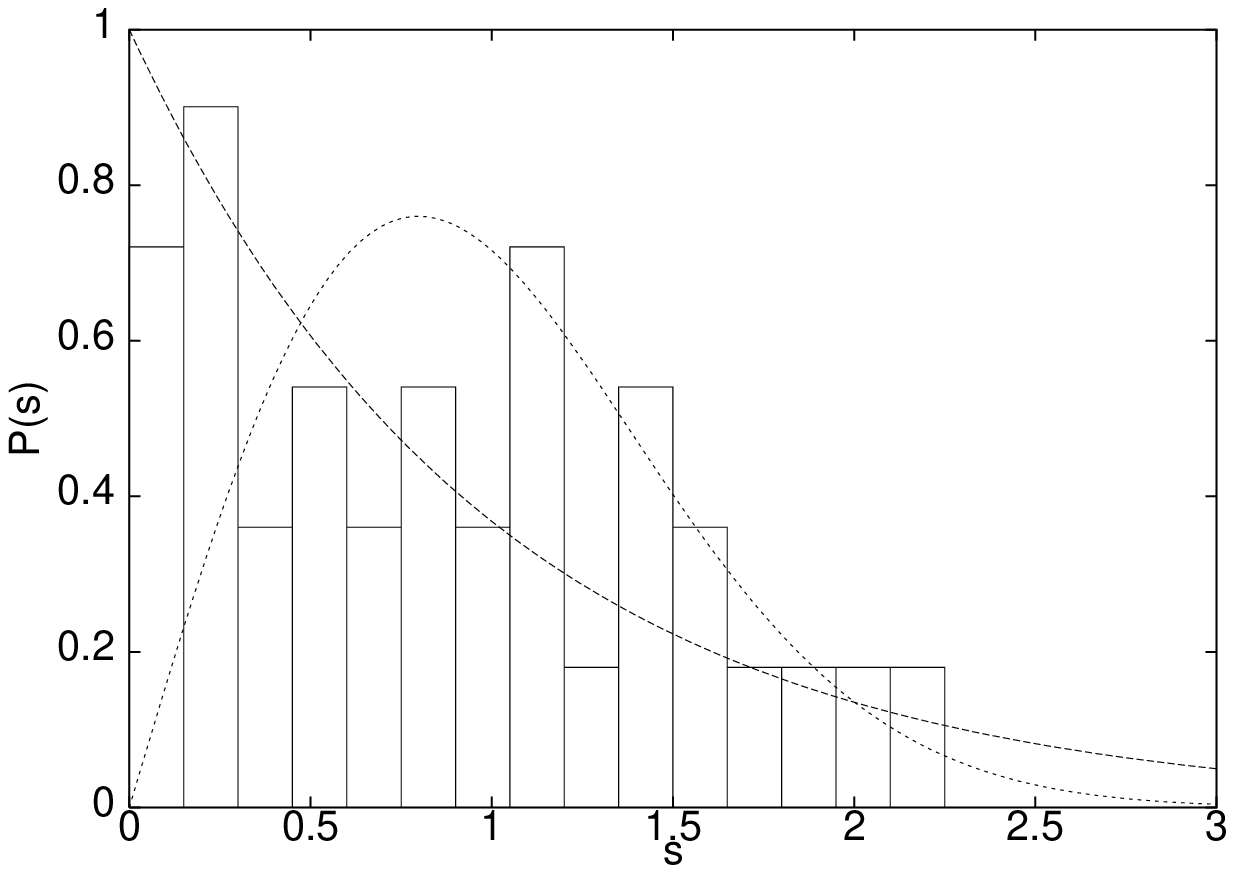,width=5cm}}}
\vspace*{0mm}
\caption{$P(s)$ for a chain with $N_s =3,5,7,9$ anharmonic oscillators
with $\lambda=1$ using a stochastic basis.
\vspace*{5mm}
\label{fig2}
 }

\end{figure*}

We computed the eigenvalues of the Hamiltonian in Eq.~(\ref{action_scalar}) 
via the Monte Carlo Hamiltonian method using a stochastic basis~\cite{Kroe}.
This is a concept to calculate eigenvalues and eigenstates (in some energy window)
of field theories or many-body systems.

To construct $P(s)$, one first has to ``unfold'' the spectra \cite{Bohi84}.
This procedure is a $local$ rescaling of the energy scale so that the mean
level spacing is equal to unity on the unfolded scale.  Ensemble and spectral
averages (the latter is possible because of the spectral ergodicity property of RMT)
are only meaningful after unfolding.

\section{Results}

Figure~\ref{fig1} shows 
the nearest-neighbor spacing distribution $P(s)$ for the
chain of harmonic oscillators with $\lambda = 0$.
This case reduces to the Klein-Gordon model which is classically
integrable and thus a free theory. The quantized system of $N_s$
oscillators has a highly degenerate spectrum. We keep in mind
that a single harmonic oscillator is an exceptional situation
which would lead to a $\delta$-function at $s=1$ after formal
unfolding. The left plot in Fig.~\ref{fig1} depicts $P(s)$
of the analytically known spectrum for $N_s=9$ oscillators. It
might consist of a washed out peak from the harmonic oscillator spacing 
and of a peak around the origin from the Poisson distribution
of the lifted degeneracies. The right plot in Fig.~\ref{fig1}
presents $P(s)$ for the corresponding spectrum obtained from
the stochastic basis (breaking parity and translational invariance)
with 1000 elements. It clearly exhibits Wigner
behavior with the underlying symmetry of the GOE. It seems
that the "holes" in the matrix produced by the method give
effectively a random matrix. Although the individual exact eigenvalues
are well reproduced~\cite{Kroe} their fluctuations are different leading
to a completely different dynamics.

Figure~\ref{fig2} shows the nearest-neighbor spacing distribution
$P(s)$ for the oscillator chain when the anharmonic term is 
switched on, $\lambda=1$. We vary the length of the chain 
from $N_s=3$ to $N_s=9$ and take $100-200$ states for the
stochastic basis (with parity symmetry implemented)
into account. The plots are inconclusive and might
show a trend to uncorrelated eigenvalues compared to $\lambda = 0$
which could be an
effect from the deeper quartic potential. It will be
interesting to see if this is enhanced with increased $\lambda$.

In summary, we have analyzed the spectrum of a finite chain of
oscillators, with and without a quartic coupling. The
infinitely long chain of harmonic oscillators is
equivalent to the Klein-Gordon model being a free theory.
Its spectrum is known analytically also for finite $N_s$.
We tested a stochastic method and obtained a spectrum
corresponding to a Wigner distribution in contrast to the
theoretical expectation. Only preliminary results could
be collected for the anharmonic oscillators and conclusions
have to be drawn in the future increasing the stochastic
basis and the coupling space.

This study is part of a general concept where random
matrix theory is used to distinguish between the quality
of a model compared to a theory or experiment. An analysis
in this respect is being performed with a quark potential
model and the experimental hadron spectrum~\cite{prep}.

{\it Acknowledgments:} 
This work was supported by NSERC Canada (H.K.), NNSF China (X.Q.L.),
and  by FWF project P14435-PHY (R.P.). We thank
G. Akemann for helpful discussions.

\end{document}